# Cassie-Wenzel transition induced by localized freezing after droplet impact on supercooled micro-patterned surfaces

Jun Fang[1], Mengqi Ye[1], Huafeng Liu[2], Yupeng Jiang[1], Tianyou Wang[1,3], Zhizhao Che[1,3,*]

1. State Key Laboratory of Engines, Tianjin University, Tianjin, 300350, China

2. National Gravitation Laboratory, MOE Key Laboratory of Fundamental Physical Quantities Measurement, Huazhong University of Science and Technology, Wuhan 430074, China

3. National Industry-Education Platform of Energy Storage, Tianjin University, Tianjin, 300350, China

*Corresponding author. Email: chezhizhao@tju.edu.cn

## Abstract

Micro-patterned surfaces have attracted significant attention in numerous applications owing to their potential to enhance hydrophobic and icephobic properties. A Cassie state of final wetting of a droplet upon impact on a micro-patterned surface, which is highly favorable for anti-icing applications, is achieved in this study through rapid localized freezing in the droplet-surface contact region via tuning the coupled interplay among droplet spreading kinetics, interfacial heat transfer, and solidification dynamics. Synchronized high-speed imaging and infrared thermography are employed to probe droplet impact and freezing dynamics, with particular emphasis on the transition of wetting state and its effect on the resulting freezing morphology. Experimental results reveal that variations in impact velocity and wall temperature lead to a final frozen wetting-state transition of the droplet from the Wenzel to the Cassie regime, accompanied by pronounced changes in freezing time, final spreading diameter, and frozen height. The transition of wetting states is attributed to rapid localized freezing at the droplet bottom, which suppresses liquid penetration into the micro-pattern. At lower impact velocities and surface temperatures, droplets tend to maintain the Cassie state with extended freezing durations, whereas higher velocities or higher temperatures promote rapid penetration and accelerated freezing. This study elucidates the coupled penetrating-freezing mechanism governed by micro-pattern design and provides fundamental insights into the rational design of anti-icing and icephobic surfaces.

**Keywords**: Cassie-Wenzel transition, Freezing, Droplet impact, Micro-patterned surface

## 1. Introduction

Droplet freezing is widely observed under supercooled conditions and is crucial for the performance and safety of various engineering systems, including aerospace [1,2], power transmission [3,4], and wind energy applications [5-7]. The freezing process during droplet impact on supercooled solid surfaces involves multiple coupled phenomena, such as droplet deformation, spreading and retraction, interfacial heat transfer, and phase-change solidification [8-11]. Complex couplings arise between droplet impact dynamics and the ensuing solidification behavior.

The icing process of droplets impacting supercooled surfaces is governed by multiple interacting parameters. Surface physicochemical characteristics strongly regulate the droplet morphology and its freezing dynamics, with wettability playing a particularly critical role. Existing studies have shown that hydrophobic surfaces markedly extend the freezing time compared with hydrophilic ones, accompanied by a 71.7% increase in droplet supercooling and a 53.9% increase in heterogeneous nucleation energy [12]. A decrease in wall temperature increases the energy dissipation rate of the droplet during impact and spreading, resulting in a shortened freezing time [13]. When droplets impact supercooled cylindrical surfaces, distinct frozen morphologies can be observed, indicating that surface curvature and the evolution of the three-phase contact line strongly regulate the solidification process [14]. Surface frosting induced by ambient humidity can also substantially modify surface characteristics, thereby influencing droplet dynamics and surface hydrophobic performance, ultimately leading to variations in droplet morphology and freezing time [15]. Furthermore, intrinsic droplet properties also have a considerable influence on droplet morphology. For instance, supercooling can reduce the maximum lateral extent reached by the droplet during spreading [13]. Increasing the impact velocity can enhance the initial kinetic energy and promote spreading, leading to a larger maximum spreading diameter, while the accelerated heat transfer and phase-change processes result in a pronounced reduction in freezing time [16].

Micro-patterned surfaces, owing to their unique capability to regulate wetting states, have emerged as an effective means to manipulate droplet behavior and control the icing process. For droplets on micro-patterned surfaces, the air pockets entrapped within the surface microtextures allow the droplet to adopt and maintain a Cassie wetting state, where an effective thermal barrier exists at the liquid-solid interface. This barrier effectively suppresses interfacial heat transfer and heterogeneous nucleation, thereby delaying the onset of freezing [17]. Furthermore, surfaces with engineered topographies can reduce droplet residence time through mechanisms such as droplet breakup [18,19], pancake-like rebound [20-22], or asymmetric recoiling [23,24], thus enhancing the surface's anti-icing performance. However, when droplets impact micro-patterned surfaces, the wetting state can undergo an irreversible transition from the Cassie to the Wenzel regime, resulting in a reduction of superhydrophobicity and loss of anti-icing functionality [25-27].

The geometric characteristics of micro-patterned surfaces, such as pillar spacing, pillar height, and groove width, play a crucial role in governing droplet spreading and wetting transitions. These structural parameters affect spreading velocity, retraction behavior, and rebound dynamics, which together determine the wetting state and energy dissipation at the liquid-solid interface [28-31]. Complex surface architectures, such as combined cylindrical-hammerhead fins and shape-memory alloy micro-patterns, have been demonstrated to actively influence droplet impact behavior and interfacial heat transfer under both high-temperature and low-temperature conditions [32,33].

Beyond influencing droplet dynamics, the geometric features of micro-patterned surfaces also significantly affect droplet solidification behavior. The duration of droplet freezing is strongly correlated with both the solid fraction and surface roughness, as well as the wetting state. In the Cassie state, higher solid fractions lead to shorter freezing times, whereas in the Wenzel state, the opposite trend can be observed [34]. Studies have shown that optimization of structural parameters such as pillar spacing, aspect ratio, and hierarchical configuration can delay the onset of droplet nucleation and the total solidification duration by 221% and 146%, respectively [35-37]. Hierarchical micro-patterned surfaces, compared with single-scale structures, exhibit superior droplet suspension and nucleation-shielding capabilities, significantly reducing the icing rate [38]. Depending on the structural scale and impact conditions, droplets can exhibit distinct freezing modes, including partial penetration, penetration with evacuation, penetration without evacuation, and internal rupture [39,40].

Although previous studies have provided valuable insights into how impact velocity, wall temperature, and micro-pattern geometry govern droplet solidification dynamics, the role of micro-patterned surfaces in regulating droplet freezing remains insufficiently understood. In this work, we discover an intriguing phenomenon in which the final stable wetting morphology of a droplet after impact and complete freezing on a micro-patterned surface is strongly affected by wall temperature, rather than being determined solely by impact dynamics. Specifically, under otherwise identical impact conditions, decreasing the wall temperature causes the final frozen wetting morphology to change from a fully penetrated Wenzel state to a partially penetrated state and eventually to a nearly non-penetrating Cassie state. Here, the wetting-state transition discussed in this work refers specifically to the variation in the final frozen wetting state across different experimental conditions, rather than to the dynamic wetting transition occurring during the impact process. Because the Cassie state is highly favorable for anti-icing applications, the finding is beneficial for numerous anti-icing applications. Analysis of droplet morphology and temperature fields indicates that this transition originates from rapid localized freezing at the droplet base, which impedes liquid penetration into the micro-pattern. Furthermore, the influence of droplet impact velocity and wall temperature on this phenomenon is systematically investigated, and a regime map of droplet impact freezing is established across a range of conditions, providing new insights into the coupled penetrating-freezing dynamics on micro-patterned surfaces.

## 2. Experimental methods

The schematic of the experimental setup is presented in Figure 1(a). A syringe pump slowly pushed deionized (DI) water from a syringe, forming droplets at the needle tip, which then fell under gravity. The initial droplet diameter was adjusted by using different needles. The impact velocity of the droplets was set by manipulating the release height, allowing precise control over the kinetic energy at impact. The temperature of the impact surface was reduced using a thermoelectric cooler, with temperature adjustments based on feedback signals. To ensure temperature uniformity on the impact surface, a copper plate was placed between the thermoelectric cooler and the surface, keeping the temperature variation within 0.1 °C. A layer of thermal paste was applied between the thermoelectric cooler and the copper plate to minimize interfacial thermal resistance. The heat generated by the thermoelectric cooler was transferred to a thermostatic bath through a water-cooled plate located at the bottom of the cooler, enabling rapid cooling of the impact surface. To prevent frost formation on the impact surface, the experimental setup was covered by an acrylic chamber, with an opening at the top to allow the droplet to enter and impact the surface. The chamber was purged with dry compressed air, maintaining the relative humidity below 10%. During the experiment, both the droplet's initial temperature and the ambient temperature were maintained at 20 ± 1 °C. A certain level of supercooling is generally required for droplet freezing. In order to promote rapid solidification after impact, the wall temperature in this study was controlled within the range of −15 °C to −27 °C. To prevent droplet breakup during impact, relatively low impact velocities (0.35–0.87 m/s) were adopted. By selecting these ranges of wall temperature and impact velocity, the coupled impact-freezing dynamics can be investigated in a controlled and systematic manner, allowing the respective influences of thermal and inertial effects on spreading, penetration, and final wetting state to be identified.

Two different types of surfaces were used in the experiment: one was a patterned silicon surface with micro-pillars formed by etching (pillar dimensions: length 0.1 mm, width 0.1 mm, height 0.5 mm, and spacing 0.1 mm), and the other was a flat silicon surface without any hydrophobic coating or additional chemical surface modification. The micro-patterned surface was fabricated directly by etching the silicon wafer, again without any hydrophobic coating or additional chemical surface modification. Their contact angles of water droplets on the surfaces are 120° and 60°, respectively, as shown in Figure 1(c).

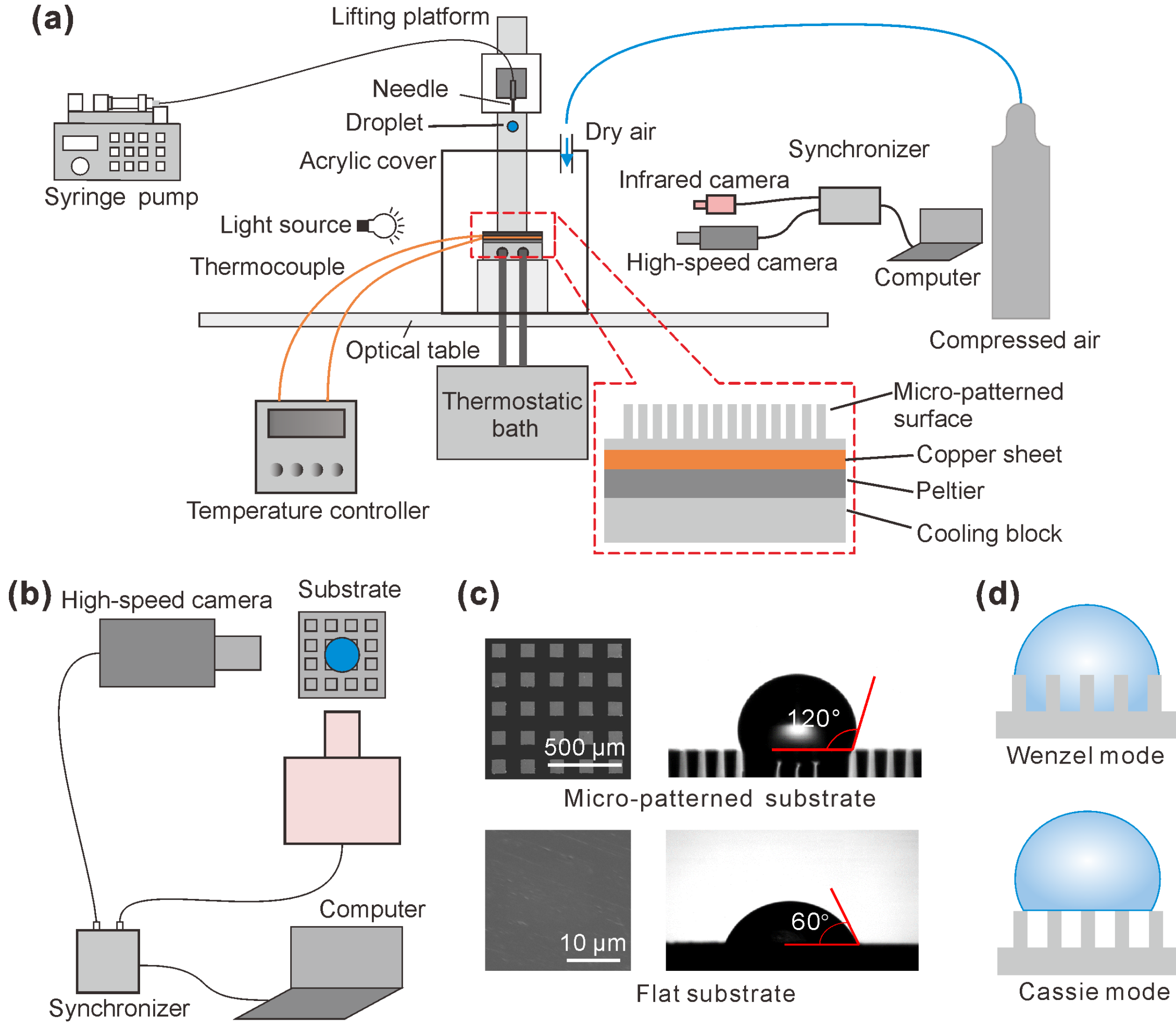

Figure 1. Experiment details of the freezing process of a droplet impacting a supercooled surface. (a) Schematic diagram of the experimental setup; (b) Schematic diagram of synchronized imaging by a high-speed camera and an infrared thermographic camera; (c) SEM images of the micro-patterned and flat surfaces and their contact angles of water droplets; (d) Schematic diagram of two different wetting states (Wenzel and Cassie) on micro-patterned surfaces.

To capture the dynamics of droplet impact, the process was recorded using a high-speed camera (Phantom T7510) equipped with a macro lens (Tokina AT-X M100 AF PRO) at 10,000 fps. LED lights and frosted glass were used as the light source. A custom digital image processing program was employed to analyze the temporal evolution of droplet morphology throughout the impact and solidification processes. To investigate the droplet's temperature distribution during the impact process, an infrared camera (InfraTec ImageIR 8355hp) was employed. The infrared camera viewed the droplet/substrate region in a direct line-of-sight configuration, without any intervening window or solid material. The high-speed camera and infrared camera were synchronized, with their imaging directions set at a 90-degree angle to each other. The high-speed camera recorded the morphological evolution of the droplet upon impact at 5000 fps, while the infrared camera recorded the temperature distribution at 250 fps. In this work, the infrared measurements were used primarily to provide a consistent comparative indication of thermal evolution and relative temperature differences among different test conditions, rather than highly precise microscale absolute temperature values. Therefore, the freezing stage discussed in this work was inferred from the combined thermal and synchronized morphological observations.

# 3. Results and discussion

## 3.1 Transition of wetting state

### 3.1.1 Transition of wetting state induced by wall temperature

Although the droplet morphology after impact is mainly determined by the impact dynamics, the freezing thermodynamics also exerts a substantial influence. In our experiments with droplets impacting micro-patterned surfaces, we found that varying the surface supercooling temperature altered the final wetting state of the fully frozen droplet. As the wall temperature decreased from −15 °C to −27 °C, the droplet's wetting state upon complete solidification transitioned from the Wenzel state to the Cassie regime, as illustrated in Figure 2. When the wall temperature was −15 °C, the droplet attained its peak spreading at 3.5 ms without penetrating the micro-pattern. After retracting to its minimum spreading diameter at 13.5 ms, it began to penetrate the micro-pattern. Due to the remaining kinetic energy, the droplet continued to oscillate, during which it slowly penetrated the surface. By 0.387 s, the droplet base fully contacted the micro-patterned surface, forming a completely penetrated Wenzel state. The droplet shape remained stable until it fully froze at 13.1 s (Figure 2d). When the wall temperature decreased to −21 °C, the droplet attained its peak spreading at 3.3 ms without penetrating the micro-pattern. After retracting to the minimum spreading diameter at 13.7 ms, it partially penetrated the micro-pattern. During the oscillation, the droplet did not continue to penetrate further, and after freezing, it exhibited a partially penetrated partial Wenzel state (Figure 2e). As the wall temperature was further reduced to −27 °C, the droplet attained its peak spreading at 3.67 ms without penetrating the micro-pattern. Although freezing had already occurred at the droplet base, the frozen region had not yet developed into a fully rigid and compact solid layer. As a result, slight further penetration could still occur in the central bottom region during the subsequent retraction stage. Throughout the retraction and oscillation, the droplet never contacted the interior of the micro-pattern. After freezing, the droplet remained in a non-penetrating Cassie state (Figure 2f). Thus, across different wall-temperature conditions, liquid penetration into the micro-pattern that would otherwise occur during impact is progressively suppressed when droplets impact colder walls, leading to substantial changes in the final stable wetting state and frozen morphology of the droplet.

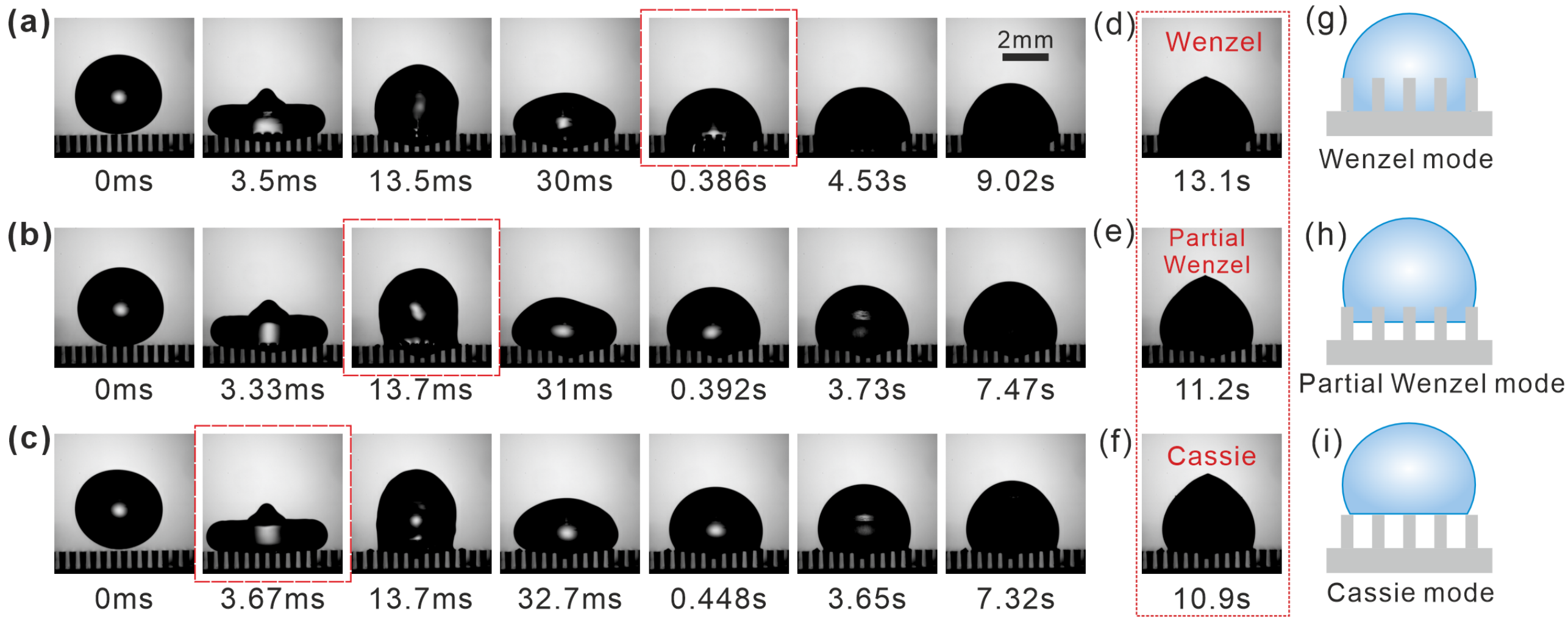

Figure 2. (a-c) Freezing of a droplet impacting a supercooled micro-patterned surface. $D_0 = 2.34$ mm, $V_0 = 0.35$ m/s, $T_s = -15$ °C (a), −21 °C (b), −27 °C (c). The red boxes in the images highlight the moments when the droplet's wetting state stops changing. Video clips are available as Supplementary Material (Videos 1–3). (d–f) Images of the droplet after completely freezing at different wall temperatures. (g–i) Schematic diagrams of the wetting states of completely frozen droplets, corresponding to panels (d–f).

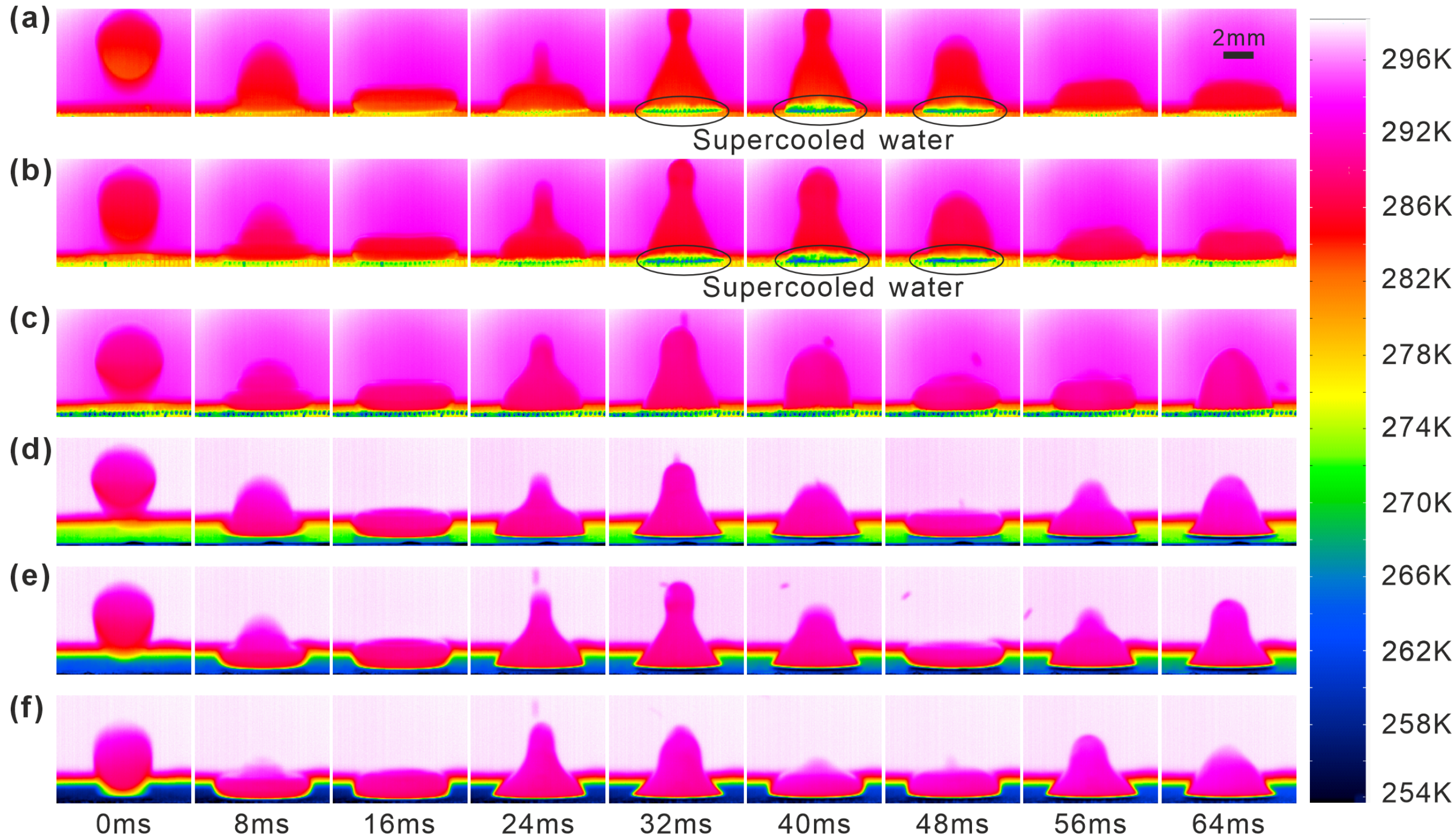

Figure 3. Infrared images of the droplet impact on supercooled surfaces: (a–c) patterned surface; (d–f) flat surface. The droplet initial diameter is $D_0 = 4.02$ mm, and the impact velocity is $V_0 = 0.35$ m/s. Wall temperatures are $T_s$ = (a) −15 °C, (b) −21 °C, (c) −27 °C, (d) −15 °C, (e) −21 °C, (f) −27 °C, respectively.

Generally, for a given micro-patterned surface, the post-impact wetting behavior of the droplet is strongly affected by the impact velocity. As the impact velocity increases, the transient impact pressure exerted on the liquid–air interface between the micro-pillars increases accordingly. When this impact pressure exceeds the capillary resistance that maintains the Cassie state, the trapped air pockets can be compressed and partially expelled, allowing the liquid to penetrate downward into the microstructure. Consequently, the droplet may undergo a Cassie-to-Wenzel wetting transition [41]. To prevent the transition of the droplet into the Wenzel wetting regime after impacting a micro-patterned surface, many methods have been tested, such as changing the solid phase fraction of the micro-pattern [42,43], adjusting the micro-pattern height [44], and applying external fields (electric or magnetic) [45,46]. In this study, we found that changing the wall temperature can also suppress the transition of the Wenzel state. Regarding the mechanism behind this phenomenon, based on the droplet morphology captured by high-speed imaging, we predict that freezing occurred before the droplet could penetrate into the micro-pattern, thereby preventing the Wenzel state and maintaining the Cassie state. To verify this hypothesis, an infrared camera was employed to measure the temperature field and monitor its dynamic evolution during the impact process. The results are presented in Figure 3.

The droplet freezing process is generally divided into five stages [47,48]: supercooling, nucleation, rewarming, freezing, and further cooling. As shown in Figure 3, when droplets impact the −15 °C and −21 °C micro-patterned surfaces, a relatively low-temperature region beneath the droplet is clearly visible in the images at 32, 40, and 48 ms. This region exhibits a lower temperature compared with the bulk droplet, and this region changes as the droplet moves, indicating the presence of flowing liquid supercooled water. At 56 and 64 ms, the temperature of the droplet’s bottom rises to near 0 °C, indicating that this region is initially supercooled and then undergoes nucleation and phase change. In contrast, when the surface temperature is −27 °C, no obvious supercooled water region is evident in the interfacial region between the droplet and surface, leading to quick freezing. This indicates that the wall temperature is too low, and the supercooling and rewarming stages are greatly shortened. Comparison of infrared images at different wall temperatures reveals that at −15 and −21 °C, a distinct supercooled water region exists at the droplet’s bottom, suggesting that the droplet has not frozen during the initial contact with the surface. However, at −27 °C, no similar phenomenon is observed, indicating that the freezing initiation time becomes earlier as the wall temperature decreases.

Based on the droplet morphology obtained from high-speed imaging and the temperature distribution recorded by infrared thermography, we propose that the rapid freezing of the droplet's bottom hinders the penetrating process. At an extremely low wall temperature (−27 °C), the droplet freezes at its base during the spreading stage, which prevents penetration into the micro-patterned surface during the subsequent retraction and oscillation phases. At a relatively low wall temperature (−21 °C), the droplet does not freeze immediately during the spreading stage, and it can penetrate into the micro-pattern during the retraction stage. However, the subsequent freezing of the bottom hinders further penetration. At a relatively high wall temperature (−15 °C), the droplet’s bottom does not freeze during either the spreading or retraction stages, allowing it to slowly penetrate into the micro-pattern during the oscillation stage, ultimately achieving full penetration of the micro-pattern. The final droplet wetting behavior under varying wall temperatures is governed by the timing of bottom freezing after impact. As the wall temperature decreases, the final wetting state progressively shifts from a fully penetrated Wenzel state to a partial Wenzel state and eventually to a nearly non-penetrated Cassie state.

To further confirm the role of the surface micro-pattern in the transition of the wetting state during the impact freezing process, we conducted experiments on flat surfaces at different wall temperatures, as shown in Figure 4. In comparison with the micro-patterned surface, the morphological differences of the droplet impacting the flat surface are more pronounced at different temperatures. As the temperature increases, the droplet contact angle at the final spreading stage becomes significantly larger, and the droplet adopts a flatter shape after freezing. Under low supercooling conditions, the droplet's shape gradually approaches a static droplet; however, under high supercooling conditions, its frozen morphology resembles that of a droplet on a micro-patterned surface at the same temperature. From Figure 3(d–f), upon impact of the droplet with the supercooled flat surface, the central bottom region of the droplet shows almost no supercooling, indicating that the droplet rapidly undergoes nucleation and freezing after contacting the substrate. Freezing proceeds more rapidly at the droplet base than on micro-patterned surfaces, indicating that droplet freezing on flat surfaces responds faster to wall temperature variations, and the overall freezing process is significantly quicker than on micro-patterned surfaces.

The differences in the droplet impact phenomena on micro-patterned and flat surfaces mainly arise from the difference in the penetration and thermal transport characteristics of the two surfaces. In the case of the flat surface, the droplet directly enters the wetting state after impact, with its bottom fully in contact with the substrate. The high-efficiency heat transfer at the interface leads to rapid freezing, resulting in significant morphological changes with the surface temperature. In contrast, the micro-patterned surface can partially maintain the droplet in a Cassie or partial Wenzel state. Air trapped between the micro-pillars at the droplet base impedes heat transfer to the substrate, thereby delaying local freezing. Thus, the droplet freezing behavior on the flat surface changes significantly with temperature, whereas the micro-patterned surface effectively mitigates this change through the regulation of the wetting state and heat transfer process.

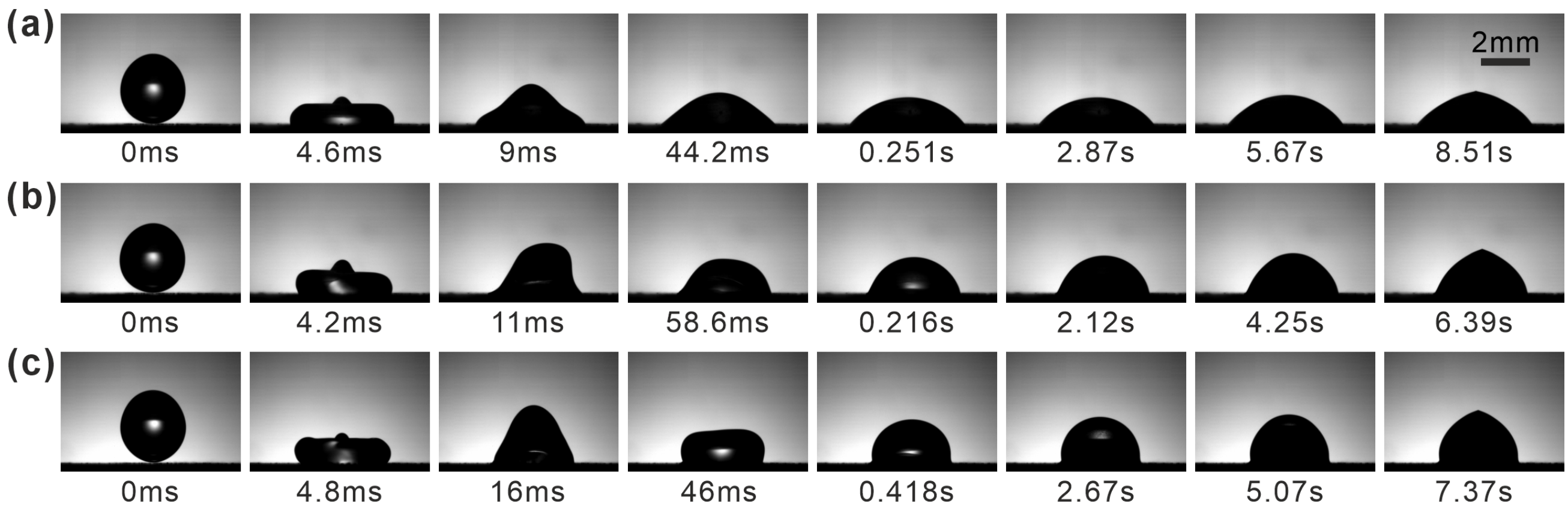

Figure 4. Freezing of a droplet impacting a supercooled flat surface. Here, the droplet initial diameter is $D_0$ = 2.34 mm and impact velocity is $V_0$ = 0.35 m/s, and the wall temperatures are $T_s$ = (a) −15 °C, (b) −21 °C, (c) −27 °C, respectively.

### 3.1.2 Transition of wetting state induced by impact velocity variation

Although low substrate temperatures speed up freezing at the droplet bottom and thereby inhibit the transition to the Wenzel wetting regime to some extent, the occurrence of bottom freezing still depends on the thermal transport process between the droplet and the substrate, which requires a certain amount of time to complete. It is therefore hypothesized that when the impact velocity of the droplet further rises, the droplet will complete the spreading and retraction in a shorter time, and its bottom does not have sufficient time to fully freeze, and the final wetting state may transition back to the Wenzel wetting regime. To evaluate the validity of this hypothesis, experiments were conducted on droplets impacting a supercooled micro-patterned surface at a range of impact velocities, as illustrated in Figure 5. Similar to the experiments at different wall temperatures, changes in the impact velocity also cause transitions in the droplet's wetting state. At a low impact velocity (0.35 m/s), the droplet attains its maximum spreading diameter at 3.9 ms but does not penetrate into the micro-pattern (Figure 5a), exhibiting a non-penetrating Cassie state. With an increased impact velocity of 0.46 m/s, the droplet begins to partially penetrate the micro-pattern at 3.5 ms when it reaches the maximum spreading diameter (Figure 5b), showing a partial Wenzel state. When the velocity further increases to 0.72 m/s, the droplet fully penetrates the micro-pattern while spreading, reaching the maximum spreading diameter at 2.6 ms, with its bottom completely penetrating the micro-pattern (Figure 5c). This results in a fully penetrated Wenzel state. The wall temperature was controlled at −27 °C throughout the experiment, corresponding to a high nucleation rate, and the droplet's bottom freezes during the spreading process. Accordingly, as the impact velocity increases, the final stable wetting state progressively shifts from a non-penetrating Cassie state to a partial Wenzel state and finally to a fully penetrated Wenzel state. Thus, it can be concluded that the penetration of the droplet into the micro-pattern during the spreading phase is a key factor determining its final wetting state. When freezing has not completed at the bottom upon contact with the substrate, it leads to penetration and the Wenzel state.

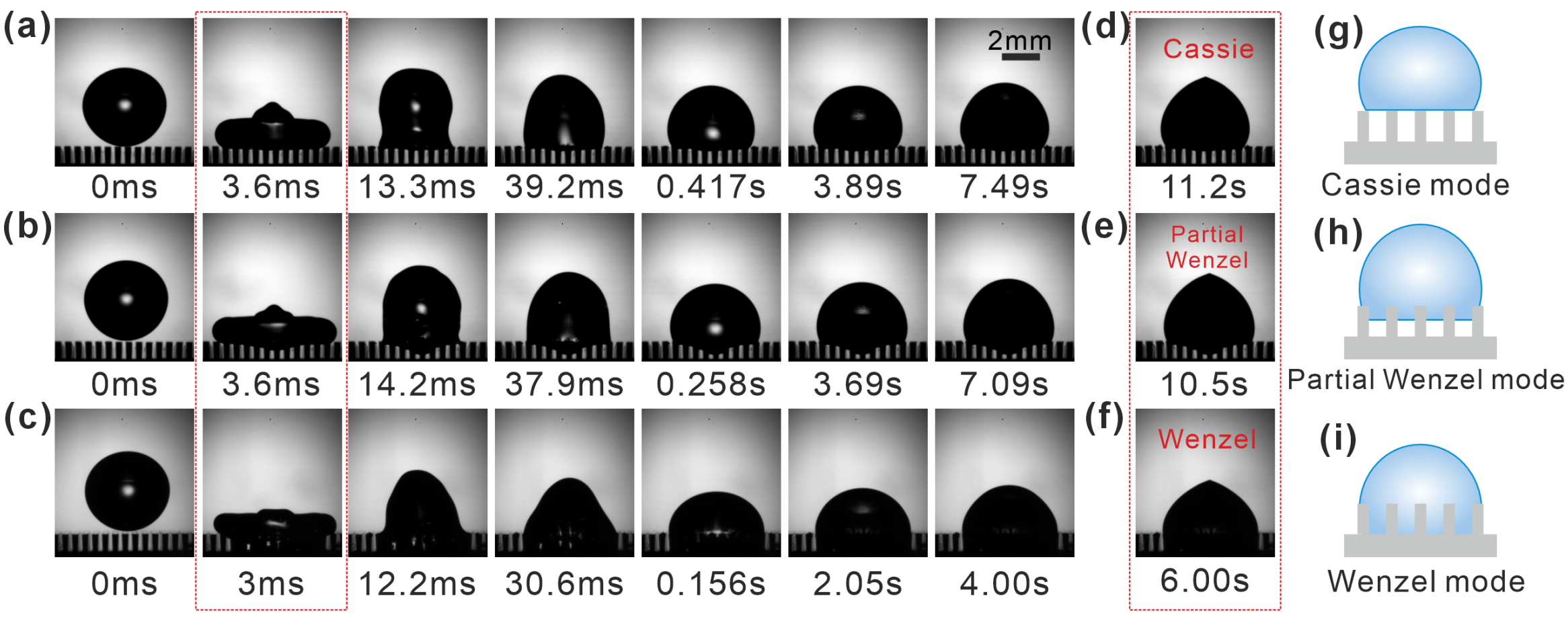

Figure 5. High-speed images of droplet freezing on a micro-patterned surface at different impact velocities. The droplet initial diameter is $D_0$ = 2.34 mm and the wall temperature is $T_s$ = −27 °C, and the impact velocities are $V_0$ = (a) 0.35 m/s, (b) 0.46 m/s, (c) 0.72 m/s. The red boxes in the images highlight the moments when the droplet's wetting state stops changing. (d–f) Snapshots of completely frozen droplets at different impact velocities. (g–i) Schematic diagrams of the wetting states of completely frozen droplets, corresponding to panels (d–f).

## 3.2 Effect of wall temperature

### 3.2.1 Freezing delay time and nucleation rate

After a droplet impacts a surface, it does not freeze immediately but undergoes a period of time before freezing begins (i.e., the freezing delay time). During this freezing delay time, supercooled water undergoes nucleation and growth. The greater the nucleation rate is, the shorter the time to freezing. The nucleation rate and freezing delay time of the droplet after impact are analyzed below. Based on classical nucleation theory, the rate of ice formation during nucleation is quantitatively described by the following expression [49]:

$$J = \Phi K(T_s) e^{-\frac{\Delta G_{\mathrm{hete}}}{K_B T_s}}, \tag{1}$$

where $\Phi$ is the solid fraction of the micro-pattern (i.e., the volume fraction of the solid phase in the micro-pattern, with $\Phi$ = 1/4 in this experiment), $K$ is the kinetic factor for the diffusion flux of water molecules at the ice-water interface, $T_s$ is the temperature of the impacted surface, and $\Delta G_{\mathrm{hete}}$ is the heterogeneous free energy barrier for the formation of the critical ice embryo at the solid-liquid interface. $K_B = 1.38\times10^{-23}$ J/K is the Boltzmann constant. The kinetic factor $K$ can be calculated using the following formula [49,50]:

$$K(T_s) = \frac{K_B T_s n}{h} e^{-\frac{\Delta F_{\mathrm{diff}}}{K_B T_s}}, \tag{2}$$

where $n = 10^{19}$ is the molecular number density of water molecules at the ice-water interface, $h = 6.63\times10^{-34}$ J·s is Planck's constant, and $\Delta F_{\mathrm{diff}}$ is the activation energy required for water molecules to diffuse through the ice-water interface. The diffusion activation energy can be calculated as [51]:

$$\Delta F_{\mathrm{diff}} = \frac{K_B E T^2}{(T - T_R)^2}, \tag{3}$$

where $E$ and $T_R$ are 892 K and 118 K, respectively [52].

Upon impacting a supercooled micro-patterned surface, the droplet undergoes heterogeneous ice nucleation at the droplet-substrate interface. According to classical nucleation theory, the free energy barrier for the formation of a critical ice embryo in heterogeneous nucleation ($\Delta G_{\mathrm{hete}}$) is smaller than that for homogeneous nucleation ($\Delta G_{\mathrm{homo}}$). The ratio between them, denoted as $f$, represents the reduction factor (where $0 < f < 1$):

$$\Delta G_{\mathrm{hete}} = f \Delta G_{\mathrm{homo}}, \tag{4}$$

$$\Delta G_{\mathrm{homo}} = \frac{16\gamma_{IW}^3}{3(\Delta G_V)^2}, \tag{5}$$

where $\gamma_{IW}$ is the ice-water interfacial tension, and $\Delta G_V$ is the volume free-energy difference between bulk ice and bulk liquid, which can be obtained using the following equations [49]:

$$\gamma_{IW} = 28.0 + 0.25(T_s - 273.15), \tag{6}$$

$$\Delta G_V = \frac{T_m - T_s}{T_m} \Delta H_V, \tag{7}$$

where $T_m = 273.15$ K is the melting temperature of ice at standard atmospheric pressure, and $\Delta H_V = 287$ MJ $m^{-3}$ is the volumetric latent heat of fusion of water.

The roughness and wetting properties of the micro-patterned surface both influence the free-energy barrier for heterogeneous nucleation. Therefore, considering their combined effects, the ratio $f$ can be expressed as [50]:

$$f = \frac{1}{2} + \frac{1}{2}\left(\frac{1-mx}{w}\right) + \frac{x^3}{2}\left[2 - 3\left(\frac{x-m}{w}\right) + \left(\frac{x-m}{w}\right)^3 + \frac{3mx^2}{2}\left(\frac{x-m}{w} - 1\right)\right], \tag{8}$$

where $m = \cos\theta_{IW}$, $x = R_n / r_c$, $w = \left(1 + x^2 - 2xm\right)$, and $R_n$ refer to the radius of the initial ice nucleus particles formed during the freezing process. $\theta_{IW}$ is the contact angle of the ice embryo in supercooled water on the nucleating particle with radius of approximately $R_n$, and $r_c$ is the critical nucleus radius, which can be expressed as [49]:

$$\cos\theta_{IW} = \frac{\gamma_I \cos\theta_I - \gamma_W \cos\theta_W}{\gamma_{IW}}, \tag{9}$$

$$r_c = \frac{2\gamma_{IW}}{\Delta G_V}. \tag{10}$$

Referring to the work of Zhang et al. [40], we use the contact angle $\theta_W$ of the droplet on the surface to represent $\theta_{IW}$. The surfaces used in the experiment were etched, unpolished silicon wafers with a contact angle of 60°. Based on Eqs. (1) – (10), a quantitative relationship between the nucleation rate $J$, impact surface temperature $T_s$, and characteristic size $R_n$ can be established. Figure 6(a) illustrates the variation of the nucleation rate $J$ with the substrate temperature $T_s$. For the regular square micropillar surface considered in the present study, the corresponding $R_n$ value is determined to be 56 μm. The detailed calculation procedure is given in the Supplementary Material S3. From the figure, with the wall temperature set at −15 °C, the nucleation rate is about $10^7$, while at −27 °C, the nucleation rate is as high as $10^{22}$, a difference of approximately 15 orders of magnitude. Figure 6(b) schematically illustrates the penetration mechanism under different degrees of supercooling. At higher supercooling, rapid localized freezing occurs near the contact region between the droplet and the micro-pattern, which suppresses further liquid penetration and leads to a non-penetrating state. In contrast, at lower supercooling, no obvious localized freezing develops at the interface, allowing the liquid to penetrate into the micro-pattern. The calculated result is consistent with the phenomenon observed in Figure 2, where the droplet rapidly freezes upon contact at −27 °C and experiences a significant delay at −15 °C.

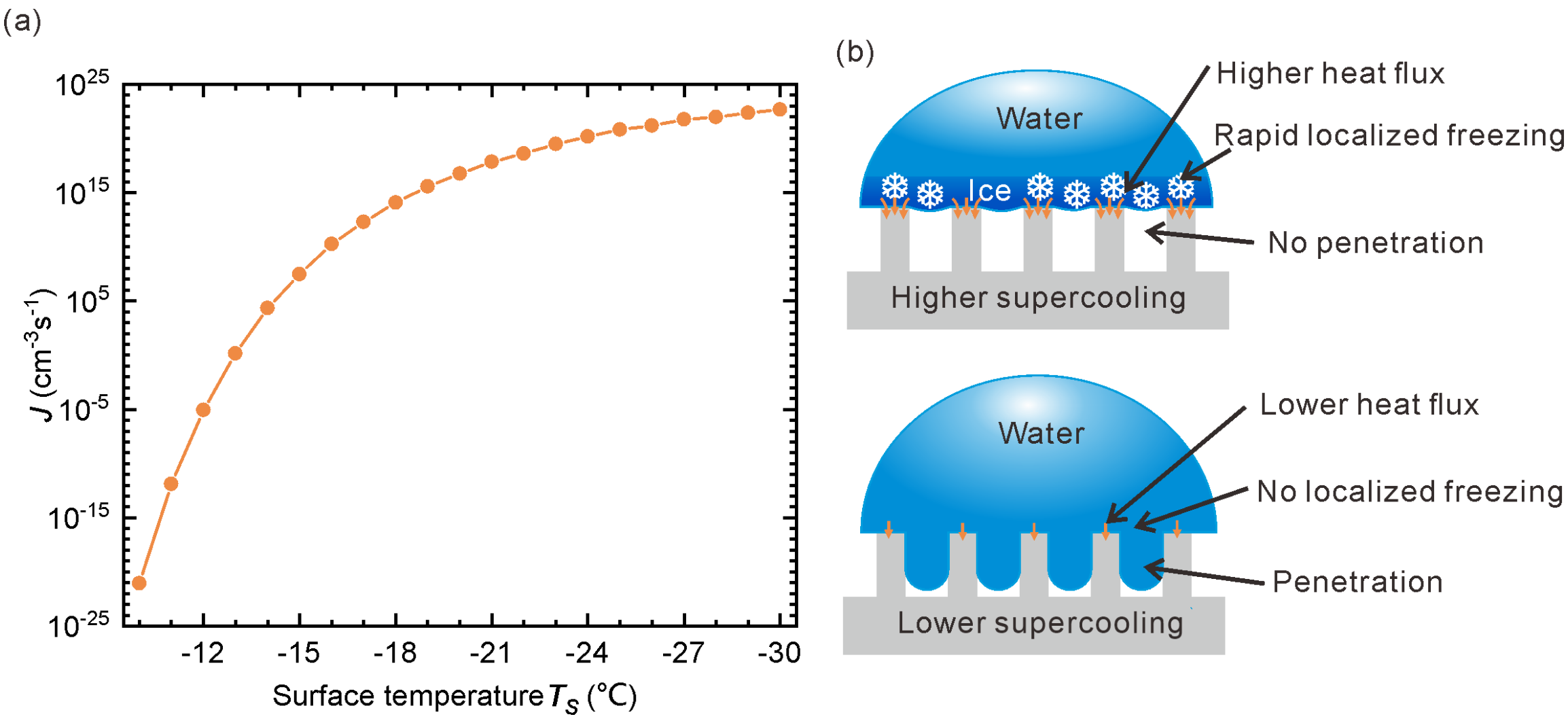

Figure 6. (a) Variation of nucleation rate $J$ with surface temperature $T_s$. (b) Schematic diagrams of droplet freezing on micropatterned surfaces at different degrees of supercooling, showing the non-penetrating state at higher supercooling degree (top) and the penetrating state at lower supercooling degree (bottom).

### 3.2.2 Droplet freezing time

Considering the critical influence of droplet freezing dynamics during impact, experiments were performed on micro-patterned surfaces at various wall temperatures to measure the droplet freezing time, which is defined as the time interval between the initial droplet-surface contact and complete solidification. For comparison under the same impact condition, the freezing time is presented in dimensionless form as $t_f^* = t_f V_0 / D_0$, where $t_f$ is the freezing time, $V_0$ is the impact velocity, and $D_0$ is the initial droplet diameter. The results demonstrate a strong correlation between the dimensionless freezing time and the final wetting state of the droplet. As shown in Figure 7(b), the freezing time does not monotonically decrease with decreasing wall temperature but instead exhibits a nonlinear trend: it first decreases, then increases, and decreases again, with two transition points observed at approximately −18 °C and −24 °C. It should be noted that the freezing time reported here represents complete solidification of the entire droplet, rather than the onset of ice nucleation or local freezing at the droplet base. Therefore, the large dimensionless freezing time does not exclude the role of early localized freezing in suppressing liquid penetration. The corresponding fully frozen images demonstrate distinct penetration behaviors under different wall temperatures. From −15 °C to −18 °C, the droplets completely penetrate the micro-pattern, exhibiting a Wenzel state. Between −19.5 °C and −24 °C, a partially penetrating transition occurs, corresponding to partial Wenzel states. At −25.5 °C and −27 °C, the droplets freeze in a nearly non-penetrating Cassie state. The findings demonstrate that the freezing duration of droplets is strongly influenced by their wetting state on micro-patterned surfaces.

To explain the influence of droplet wetting state on freezing time, images of fully frozen droplets were analyzed using a digital image processing routine (8.2 μm/pixel) to measure the total projected droplet area $A_d$ and the projected droplet area penetrating into the micro-patterned region $A_c$. The projected penetration-area fraction was defined as $A_c/A_d$. Figure 7(b) shows that the projected penetration-area fraction decreases continuously with decreasing wall temperature, indicating reduced contact between the droplet and the micro-patterned surface. It decreases from approximately 10% at −15 °C to less than 1.5% at −27 °C, corresponding to a reduction to nearly 1/7 of the value at −15 °C. In the experiment, both the droplet's initial temperature and the ambient temperature were maintained at 20 ± 1 °C. Thus, a certain degree of supercooling was generated when the droplet came into contact with the cold substrate. To rationalize the observed freezing-time trends, we decompose the total freezing time as $t_f = t_n + t_g$, where $t_n$ is the nucleation waiting time (from initial contact to the first detectable nucleation) and $t_g$ is the post-nucleation freezing-growth time (from the first detectable nucleation to complete solidification). As the wall temperature decreases, the nucleation rate increases, so $t_n$ generally decreases, which explains the observed decrease in freezing time at wall temperatures between −15 °C and −18 °C as well as between −24 °C and −27 °C. However, variations in the moment of nucleation upon impact lead to distinct penetration transitions and thus different interfacial contact conditions. Meanwhile, the freezing-growth time can be estimated by a heat-transfer-limited scaling

$$t_g \sim \frac{Q}{h_{eff} A_{eff} \left(T_m - T_s\right)} \tag{11}$$

where $Q$ is the heat that must be removed (dominated by latent heat during solidification), $A_{eff}$ is the effective solid-liquid thermal contact area, and $h_{eff}$ is an effective heat-transfer coefficient incorporating interfacial thermal resistance and the micropatterns. As the wall temperature decreases, the wetting state shifts from Wenzel to Cassie, which reduces $A_{eff}$ and increases the effective interfacial thermal resistance, thereby increasing $t_g$, even though the driving temperature difference ($T_m - T_s$) becomes larger. Furthermore, lower wall temperatures lead to larger apparent contact angles after freezing, resulting in an increased droplet height above the micro-pattern and a longer propagation path of the freezing front, which further increases $t_g$. Consequently, within the intermediate wall temperature interval of −18 °C to −24 °C, the increase in $t_g$ outweighs the decrease in $t_n$, causing the overall freezing time and thus also the dimensionless freezing time $t_f^*$ to increase with decreasing wall temperature. Overall, the droplet freezing time exhibits a non-monotonic dependence on wall temperature, characterized by a decrease-increase-decrease trend.

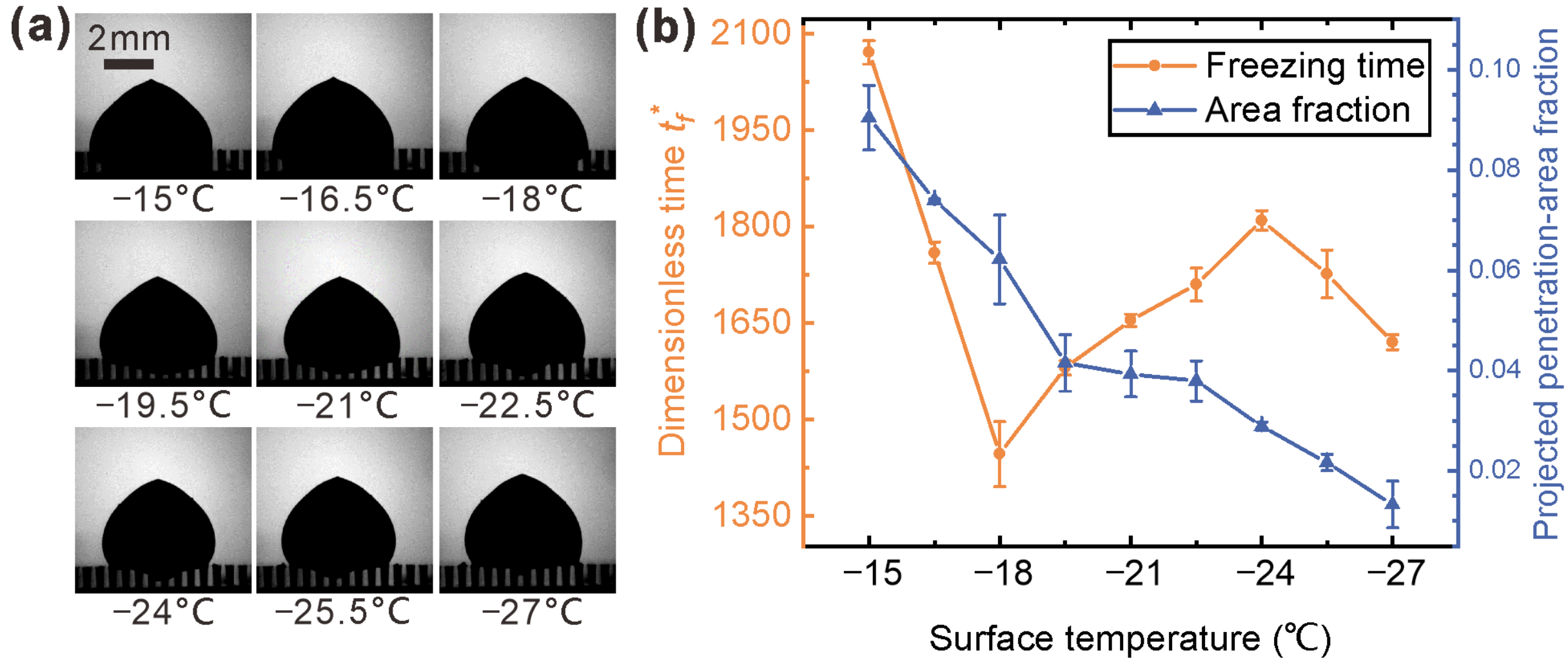

Figure 7. (a) Images of droplets completely frozen after impacting a micro-patterned surface at different wall temperatures. Here, the droplet initial diameter is $D_0$ = 2.34 mm and the impact velocity is $V_0$ = 0.35 m/s. (b) Variation of the dimensionless freezing time and projected penetration-area fraction (i.e., ratio of the droplet area penetrating into the micro-patterns to the total droplet area in the images) for droplets impacting the micro-patterned surface under different wall temperatures. The dimensionless freezing time is defined as $t_f^* = t_f V_0 / D_0$, where $t_f$ is the freezing time, $V_0$ is the impact velocity, and $D_0$ is the initial droplet diameter. Each data point represents the mean value of repeated experiments ($N \geq 3$), and the error bars denote the standard deviation. Unless otherwise specified, this definition of error bars applies to all figures throughout the manuscript.

### 3.2.3 Comparison of morphological and thermal evolution between micro-patterned and flat surfaces

To investigate the effect of surface micro-patterns on droplet impact freezing, we compared the freezing processes on micro-patterned and flat surfaces. For droplets impacting the flat surface, distinct differences in droplet morphology are observed at the moment of maximum spreading under different wall temperatures. At wall temperatures between −15 °C and −18 °C, the droplet attains its maximum spreading diameter ($D_{S\text{-}max}$) near the droplet bottom; however, as the wall temperature decreases below −19.5 °C, $D_{S\text{-}max}$ shifts toward the middle of the droplet, as shown in Figure 8(a). Compared with the micro-patterned surface (Figure 2), the morphological variations during spreading on the flat surface are more pronounced. These differences during spreading also result in significant variations in the final frozen morphology. As shown in the lower panels of Figure 8(a), droplets on the flat surface exhibit larger changes in both apparent contact angle and frozen height (considering only the portion of the droplet above the micro-patterns for comparison) with decreasing wall temperature, whereas the variations on the micro-patterned surface (Figure 7) are relatively moderate.

Based on the qualitative trends described above, we further analyzed droplet images captured at maximum spreading and complete freezing on both micro-patterned and flat surfaces, obtaining the quantitative relationships between the spreading factor $\beta_S$, the contact spreading factor $\beta_C$, and wall temperature for the two surface types, as illustrated in Figures 8(b) and 8(c). Here, $\beta_S = D_S / D_0$ and $\beta_C = D_C / D_0$ , where $D_S$ is the maximum horizontal diameter during droplet spreading and $D_C$ is the diameter of the droplet-surface contact region. Overall, droplet morphology on the flat surface is highly sensitive to wall temperature. By contrast, on the micro-patterned surface, both the spreading and contact factors at maximum spreading ($\beta_{S\text{-}max}$ and $\beta_{C\text{-}max}$) are relatively insensitive to wall temperature; however, the frozen values ($\beta_{S\text{-}frozen}$ and $\beta_{C\text{-}frozen}$) exhibit a non-monotonic variation with decreasing temperature, characterized by an initial decrease, a subsequent increase, and a final decrease. In contrast, on the flat surface, $\beta_S$ and $\beta_C$ generally decrease first and then increase as the wall temperature decreases.

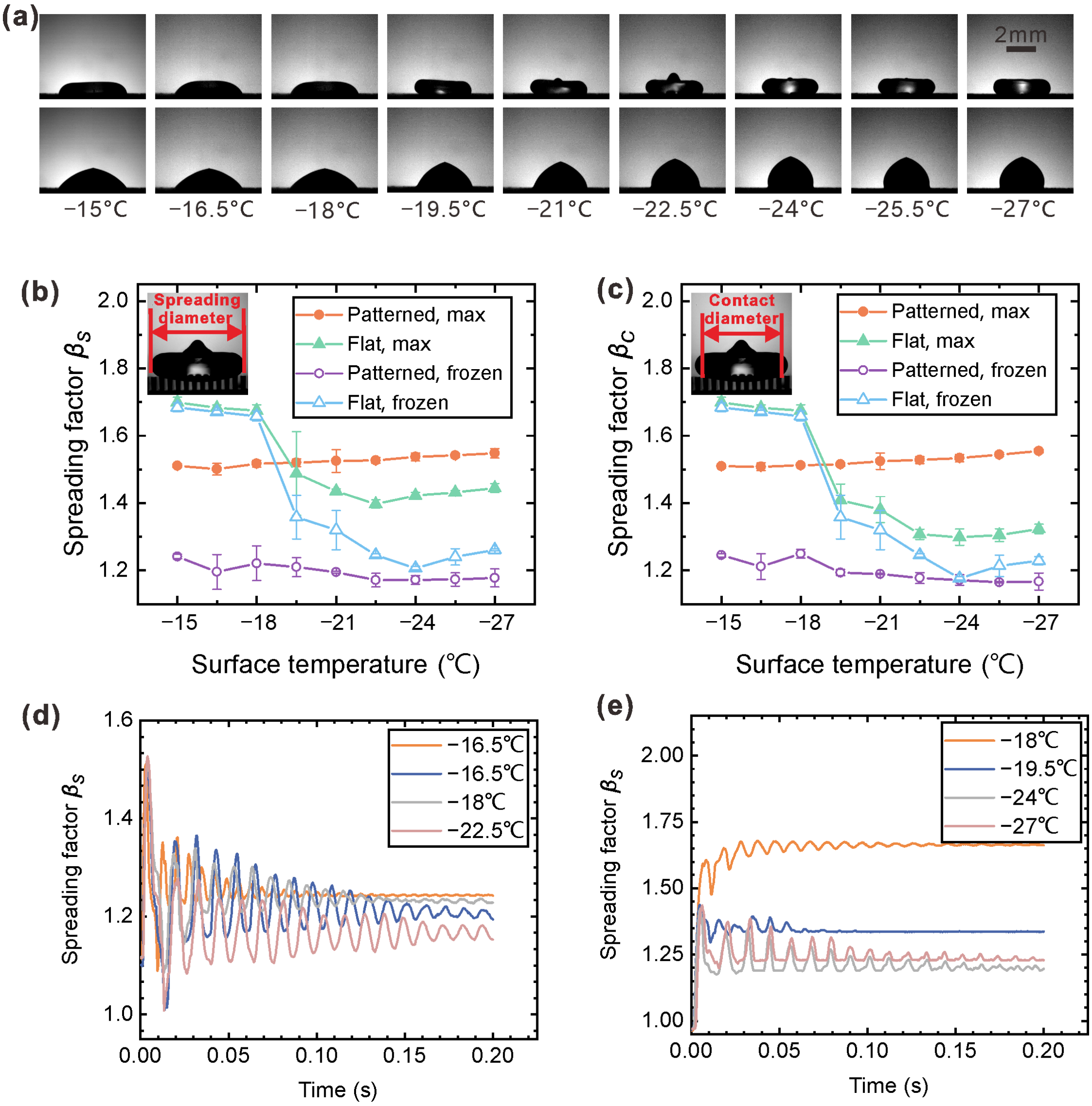

Figure 8. (a) High-speed images of droplets impacting the flat surface at different wall temperatures, showing the moment of maximum spreading (top) and the fully frozen state (bottom). Here, the droplet initial diameter is $D_0 = 2.34$ mm and the impact velocity is $V_0 = 0.35$ m/s. (b) Maximum and fully frozen spreading factors $\beta_{S\text{-}max}$ and $\beta_{S\text{-}frozen}$ for droplets impacting the micro-patterned and flat surfaces at different wall temperatures, where $\beta_S = D_S / D_0$ and the spreading diameter $D_S$ denotes the maximum horizontal diameter during droplet spreading, as illustrated in the inset. (c) Maximum and fully frozen contact spreading factors $\beta_{C\text{-}max}$ and $\beta_{C\text{-}frozen}$ for droplets impacting the micro-patterned and flat surfaces at different wall temperatures, where $\beta_C = D_C / D_0$ . The contact diameter $D_C$ denotes the diameter of the droplet-surface contact region, as illustrated in the inset. (d) Temporal evolution of the spreading factor $\beta_S$ on the micro-patterned surface at different wall temperatures. (e) Temporal evolution of the spreading factor $\beta_S$ on the flat surface at different wall temperatures.

The differences observed between micro-patterned and flat surfaces at various temperatures primarily arise from the variations in penetration behavior and freezing dynamics. On micro-patterned surfaces at different temperatures, droplets do not penetrate into the micro-patterns during the spreading stage. Consequently, the micro-patterns effectively diminish the droplet-substrate interfacial contact area, rendering the $\beta_{S\text{-}max}$ and $\beta_{C\text{-}max}$ relatively insensitive to wall temperature (Figure 8d). When droplets impact supercooled micro-patterned surfaces at −15 to −18 °C, partial penetration occurs gradually during the retraction and oscillation phases, ultimately leading to complete penetration. Notably, the minimal $\beta_S$ during oscillation cycles is smallest at intermediate temperatures because droplets at −15 °C and −18 °C quickly reach equilibrium. Although $\beta_{S\text{-}max}$ during the initial oscillation cycles shows minimal difference, it gradually decreases with continued oscillations, ultimately resulting in reduced $\beta_{S\text{-}frozen}$ and $\beta_{C\text{-}frozen}$. When the wall temperature drops below −19.5 °C, droplets are unable to fully penetrate into the micro-patterns, and oscillation duration is significantly extended, leading to negligible variation in $\beta_{S\text{-}frozen}$ and $\beta_{C\text{-}frozen}$ with temperature. In contrast, on flat surfaces, different trends are observed. At wall temperatures of −15 to −18 °C, both the $\beta_{S\text{-}max}$ and $\beta_{C\text{-}max}$ are relatively large. This is attributed to the delayed freezing process: upon first reaching the maximum spreading diameter, the droplet edges remain unfrozen while the center has already solidified, allowing the remaining kinetic energy to continue driving edge spreading (Figure 8e). When the wall temperature decreases to −19.5 °C or below, the droplet spreading during oscillation does not exceed the initial maximum, and $\beta_S$ and $\beta_C$ decrease significantly within the range of −18 °C to −19.5 °C. Further reduction of the wall temperature below −24 °C leads to increased solid-liquid contact at the droplet-substrate interface, enabling full release of inertial energy and an increase in $\beta_{S\text{-}max}$ and $\beta_{C\text{-}max}$ with decreasing temperature. $\beta_{S\text{-}frozen}$ and $\beta_{C\text{-}frozen}$ follow the same trends observed at the moment of maximum spreading. This trend can be further supported by the temperature evolution within the droplet, with a further description provided in S1 of the Supporting Information.

### 3.3 Effect of droplet impact velocity

#### 3.3.1 Droplet morphology evolution

Among the factors affecting droplet impact, the impact condition characterized by the Weber number (*We*) plays a critical role. Here, $We = \rho U_0^2 L / \sigma$ , where $\rho$ is the liquid density, $U_0$ is the impact velocity, $D_0$ is the initial droplet diameter, and σ is the surface tension. Figure 9(a) shows images of droplets impacting supercooled micro-patterned surfaces at different *We*, capturing both the moment of maximum spreading and the completely frozen state. As *We* increases, droplets increasingly penetrate the micro-patterns during maximum spreading, resulting in a transition of the wetting state from the nearly non-penetrating Cassie state to the fully penetrating Wenzel state. $D_{S\text{-}max}$ shows only minor changes, whereas the $D_{S\text{-}frozen}$ decreases noticeably. Figure 9(b) presents the relationships between *We* and $\beta_{S\text{-}max}$, $\beta_{S\text{-}frozen}$, and the normalized frozen droplet height $\alpha$ (defined as $\alpha = H / D_0$), while Figure 9(c) shows the projected penetration-area fraction of the droplet (i.e., the ratio of the droplet area within the micro-patterns to the overall droplet area, as quantified by image analysis) plotted against *We*. As *We* increases, the corresponding kinetic energy of the droplet also rises proportionally, which tends to increase $\beta_{S\text{-}max}$. However, the increased penetration into the micro-patterns reduces the non-penetrated volume of the droplet and dissipates kinetic energy more rapidly, which tends to decrease $\beta_{S\text{-}max}$. The combined effects result in only a minor change in $\beta_{S\text{-}max}$ with increasing *We*. For the frozen morphology, at relatively low *We* (3.95–6.91), an increase in *We* results in a negligible change in the frozen contact spreading factor $\beta_{C\text{-}frozen}$ but a noticeable decrease in the contact angle. Partial penetration into the micro-patterns reduces the non-penetrated volume of the droplet, slightly decreasing $\beta_{S\text{-}frozen}$. As *We* increases further, the frozen contact angle continues to decrease, $\beta_{C\text{-}frozen}$ increases, and the penetrated droplet volume grows, collectively lowering $\alpha_{frozen}$. Therefore, as *We* increases, the wetting state on the micro-patterned surface transitions from Cassie to Wenzel, and the droplet spreading and frozen morphology result from the combined effects of kinetic energy dissipation and micro-pattern penetration.

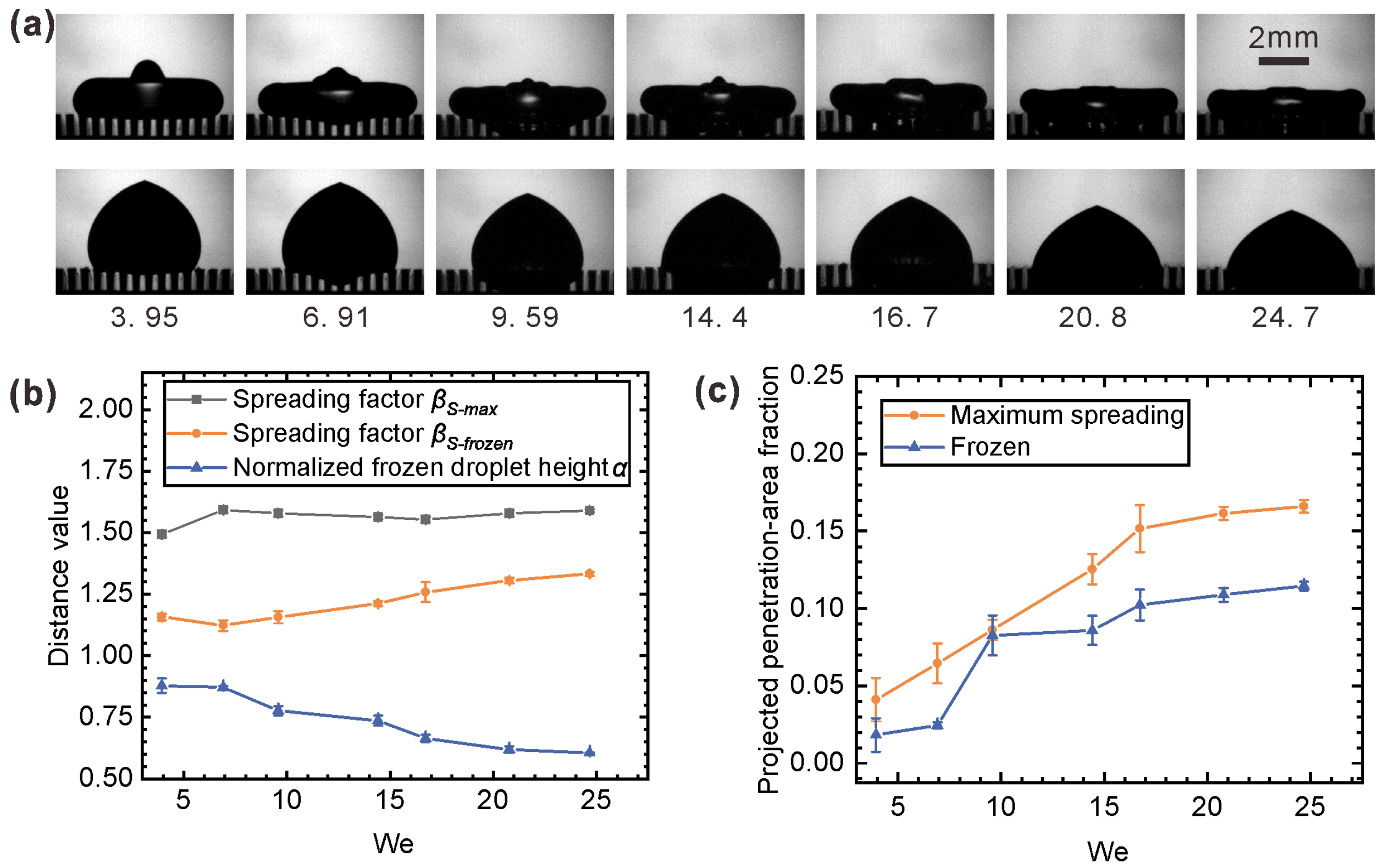

Figure 9. Morphology of droplet spreading and freezing at different Weber numbers (*We*). (a) Images of droplets impacting supercooled micro-patterned surfaces at different *We*, showing the moment of maximum spreading (top row) and the fully frozen state (bottom row). (b) Maximum spreading factor $\beta_{S\text{-}max}$, frozen spreading factor $\beta_{S\text{-}frozen}$, and normalized frozen droplet height $\alpha$ as functions of *We* on micro-patterned surfaces, where $\alpha = H_{frozen} / D_0$. (c) Projected penetration-area fraction (i.e., ratio of the droplet area penetrating into the micro-patterns to the total droplet area in the images) as a function of *We*.

### 3.3.2 Comparison of droplet morphology on micro-patterned and flat surfaces

To compare the dynamics of droplet spreading and solidification on micro-patterned and flat surfaces, we conducted additional experiments with droplets impacting a supercooled flat surface at various Weber numbers (*We*). Images of the droplets at their maximum spreading state and after complete freezing are shown in Figure 10(a). Analysis of images at varying *We* shows that $D_{S\text{-}max}$ increases with *We*, while $H_{frozen}$ decreases with *We*, showing consistent trends. At all *We*, droplets on micro-patterned surfaces generally have smaller dimensions ($D_{S\text{-}max}$, $D_{C\text{-}max}$, $D_{S\text{-}frozen}$, and $D_{C\text{-}frozen}$) than those on flat surfaces. This is partly because partial penetration into the micro-patterns reduces the droplet volume above the structures, and partly because the hydrophobicity of the micro-patterned surface increases the contact angle, thereby inhibiting spreading, whereas the more hydrophilic flat surface promotes spreading due to its lower contact angle. The maximum spreading factor of an impacting droplet is governed by the combined effects of inertia, capillarity, and viscous dissipation, and is therefore generally described as a function of both the Weber number and the Reynolds number. The maximum spreading factor can be expressed as [53]:

$$\beta_{\max} Re^{-1/5} = \frac{P^{1/2}}{A + P^{1/2}} \tag{12}$$

$$P = We Re^{-2/5} \tag{13}$$

The Reynolds number can be expressed as a function of the Weber number for pure water droplets with a fixed initial diameter:

$$Re = \frac{\sqrt{\rho \sigma D_0}}{\mu} We^{1/2} \tag{14}$$

For the present pure water droplets with $D_0 = 2.34$ mm, the physical properties of water are taken as $\rho$ = 997 kg m$^{-3}$, $\sigma$ = 0.072 N m$^{-1}$, and $\mu = 1.0\times10^{-3}$ Pa·s. With $A \approx 1.24$ [53], the relation between maximum spreading factor and Weber number can be reduced to

$$\beta_{\max} = \frac{We^{1/2}}{1.24 + 0.3We^{2/5}} \tag{15}$$

as indicated in Figure 10(b), where the experimental data remain in good agreement with the theoretical prediction under the present conditions. Once *We* increases to above approximately 20.8, this difference becomes pronounced: on the micro-patterned surface, $\beta_{S\text{-}max}$ and $\beta_{C\text{-}max}$ are reduced by about 20% compared with the flat surface, while $\beta_{S\text{-}frozen}$ and $\beta_{C\text{-}frozen}$ decrease by more than 30%. However, at lower *We*, $\beta_{S\text{-}max}$ on the micro-patterned surface can be larger than on the flat surface (Figure 10b). This occurs because the droplet penetration into the micro-patterns is minimal, leaving the volume above the micro-patterns almost unchanged. Furthermore, the gaps between micro-patterns reduce the actual contact area and heat transfer rate, hence freezing has less influence on the spreading process, resulting in a larger $\beta_{S\text{-}max}$ on the micro-patterned surface than on the flat surface.

By comparing the relationships between droplet geometric parameters and *We* on supercooled micro-patterned and flat surfaces (Figure 10b-e), distinct differences in the trends between $\beta_S$ and $\beta_C$ of droplets on micro-patterned surfaces are observed. Specifically, with the increase of *We* from 14.4 to 16.7, $\beta_{S\text{-}max}$ and $\beta_{C\text{-}max}$ exhibit opposite trends. $\beta_{C\text{-}max}$ increases with *We*, whereas $\beta_{S\text{-}max}$ decreases. Referring to the images corresponding to maximum spreading in Figure 9(a), this occurs because an increasing portion of the droplet penetrates into the micro-patterns, thereby reducing the liquid volume above the micro-patterns. Therefore, although $\beta_{C\text{-}max}$ increases, $\beta_{S\text{-}max}$ decreases. Furthermore, after completely freezing, $\beta_{C\text{-}frozen}$ continues to increase with *We*. However, the accompanying reduction in contact angle leads to a flatter droplet shape, causing $\beta_{S\text{-}frozen}$ to decrease at lower *We*. Therefore, with the increase of *We* from 3.95 to 6.91, $\beta_{C\text{-}frozen}$ increases with *We*, while $\beta_{S\text{-}frozen}$ shows a decreasing trend.

During the transition of the droplet wetting state, pronounced changes were observed in the apparent contact angle of the droplet at the apex of the micro-patterned pillars: when the state transitions from Cassie to Wenzel, the contact angle decreases from obtuse to acute, indicating a loss of hydrophobicity of the micro-patterned surface, as shown in Figure 9(a). In the process of droplet impact and subsequent freezing on a supercooled surface, the spreading stage is very short, approximately 3–4 ms, and is primarily governed by the droplet's kinetic energy. Figure 2 shows that the droplet images at the maximum spreading state under different substrate temperatures are very similar, indicating that the freezing process exerts a negligible influence on the extent of droplet penetration into the micro-patterns during spreading. Therefore, the droplet's penetration into the micro-patterned surface upon impact at initial velocity $U_0$ can be simplified as a surface-tension-dominated uniform deceleration motion [39], and the acceleration of the liquid penetration process into the micro-patterns can be expressed as:

$$a_c = -\frac{F_c}{\rho A d_{\text{film}}(1-\varphi)} = -\frac{4\varphi\sigma\left|\cos\theta_a\right|}{\rho P d_{\text{film}}(1-\varphi)}, \tag{16}$$

where $F_c$ is the surface tension force, $A$ is the projected area of the liquid film, $\theta_a$ is the contact angle on the micro-patterned surface, and the penetration depth of the liquid into the micro-patterns can be approximated as $d_{\text{film}} \sim D_0 We^{-1/2}$ [40,54]. The maximum penetration depth of the droplet can be expressed as:

$$H_i = \frac{U_0^2}{2a_c} \sim \frac{We^{1/2} P(1-\varphi)}{\varphi\left|\cos\theta_a\right|}. \tag{17}$$

When the penetration depth $H_i$ is smaller than the height of the micro-pattern pillars $H$, the droplet does not fully penetrate the micro-patterned surface. Therefore, the dimensionless penetration depth $h_i = H_i / H$ can be derived as [40]:

$$h_i = \frac{\sqrt{We}}{8\left|\cos\theta_a\right|}\frac{P}{H}\frac{1-\varphi}{\varphi}. \tag{18}$$

Here, the ratio between the micro-pillar side length and its height $P/H$ and the solid fraction of the micro-pattern $\varphi$ are the key parameters influencing the droplet penetration process. Considering the specific surface parameters adopted in this study, the transition of the wetting state can be achieved at an impact velocity scale of approximately 1 m/s, which is in good agreement with the experimental observations reported herein.

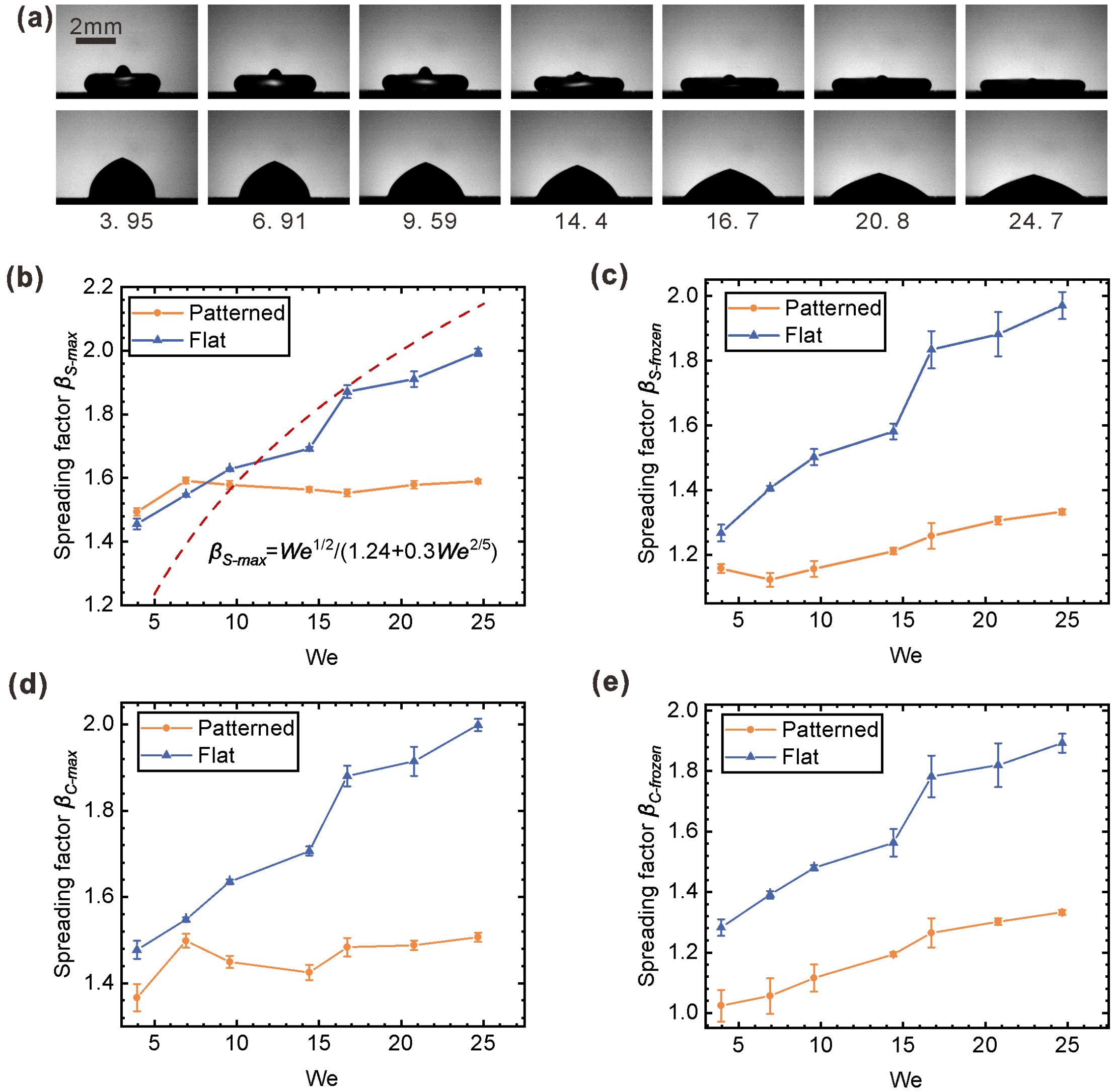

Figure 10. Comparison of morphologies between patterned and flat surfaces at different Weber numbers (*We*). (a) Images of droplets impacting flat surfaces at different *We*, showing the moment of maximum spreading (top row) and the fully frozen state (bottom row). (b–e) Relationships between spreading factors and *We* on micro-patterned and flat surfaces: (b) maximum spreading factor $\beta_{S\text{-}max}$, where the red dashed line indicates Eq. (15), which is derived from the relationship between the maximum spreading factor $\beta_{S\text{-}max}$, Weber number (*We*) and Reynolds number (*Re*) [53], (c) contact spreading factor at maximum spreading $\beta_{C\text{-}max}$, (d) fully frozen spreading factor $\beta_{S\text{-}frozen}$, and (e) fully frozen contact spreading factor $\beta_{C\text{-}frozen}$.

### 3.4 Phase diagram of droplet wetting state

Under different substrate temperatures and impact velocities, we conducted extensive experiments on droplet impacts onto supercooled micro-patterned surfaces. Fully frozen images were used to assess droplet wetting states on micro-patterned surfaces, based on which a phase diagram of droplet wetting states was constructed, as shown in Figure 11. Here, the impact condition is characterized by the Weber number $We = \rho U_0^2 L / \sigma$ , and the substrate temperature is represented by the Stefan number $St = C_p \left(T_m - T_s\right) / L$ , which is the ratio of sensible heat to latent heat. Here, $C_p$ is the specific heat capacity of liquid water at constant pressure ($C_p \approx 4.18\times10^3$ J·kg$^{-1}$·K$^{-1}$), $L$ is the latent heat of fusion of water ($L \approx 3.34\times10^5$ J·kg$^{-1}$), $T_m$ is the melting point (0 °C for water), and $T_s$ denotes the substrate temperature. The results show that droplets exhibit a Cassie state only at low *We* and high *St* (i.e., colder substrates). With increasing *We* or decreasing *St*, droplets enter a partial-penetration partial Wenzel state, and at sufficiently high *We* or low *St*, they further transition to a full-penetration Wenzel state. With decreasing substrate temperature, the droplet's freezing delay time is significantly shortened, causing the droplet's bottom to freeze rapidly during the spreading stage, which inhibits further penetration into the micro-patterned surface. However, the transfer of heat between the droplet and substrate still requires a certain time. When the impact velocity increases, the enhanced inertia and impact pressure enable the droplet to overcome the support of air pockets between microstructures and penetrate into the micro-patterned surface during spreading. Therefore, the competition between heat transfer and fluid dynamics governs the droplet's final wetting state, leading to its dependence on *We* and *St*.

Droplet size effects are also examined, and it is found that variations in the initial droplet diameter have little impact on wetting state transitions. The details are provided in S2 in Supporting Information.

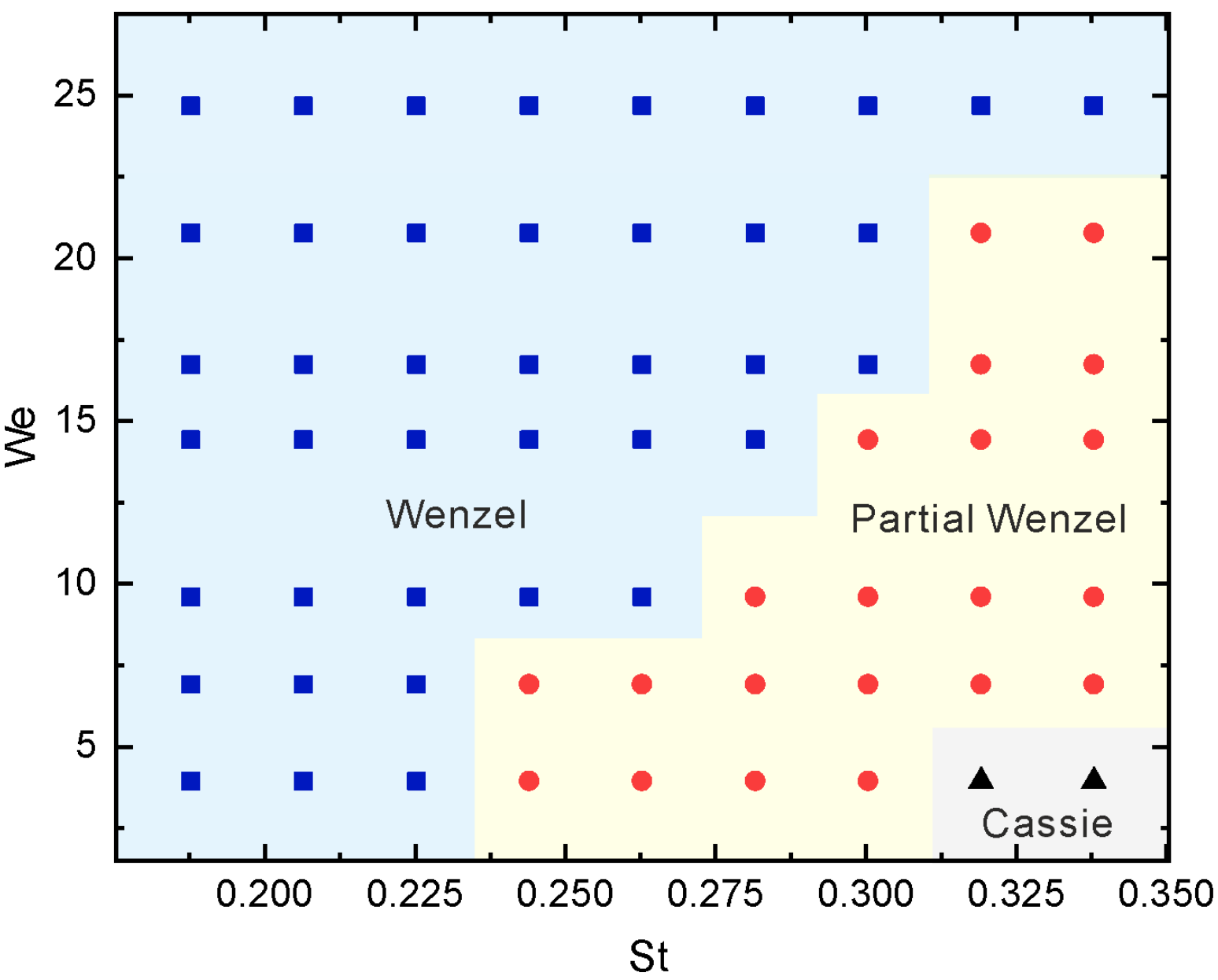

Figure 11. Phase diagram of the droplet wetting states after completely freezing on supercooled micro-patterned surfaces at different Stefan numbers (*St*) and Weber numbers (*We*).

## 4. Conclusions

In this study, we discover that a non-penetrating Cassie state can be preserved after droplet impact and complete freezing on a micro-patterned surface through rapid localized freezing at the droplet–substrate interface. The novelty of this work lies in identifying and quantifying a temperature-controlled wetting/penetration mechanism during droplet impact on micro-patterned surfaces. We study droplet impact on supercooled micro-patterned surfaces, simultaneously recording the impact, spreading, and freezing processes as well as temperature evolution using synchronized high-speed imaging and infrared thermography. Detailed comparisons are performed with flat surfaces. The results indicate that wall temperature plays a decisive role in controlling droplet nucleation and the associated wetting behavior. As the wall temperature decreases, the nucleation rate of droplets increases evidently, reaching $10^{22}$ times higher at −27 °C than at −15 °C, which leads to earlier freezing and limited penetration into the micro-patterns during the spreading and oscillation stages. Consequently, the final wetting state transitions progressively from the fully penetrated Wenzel state to the fully non-penetrating Cassie state. This is different from most previous studies, in which the Cassie-Wenzel transition is mainly tuned by increasing impact inertia (e.g., impact velocity/Weber number) or by applying external fields (e.g., electric or magnetic fields). Here, we demonstrate that wall temperature itself can act as an effective control parameter: rapid freezing at the droplet base suppresses penetration and preserves the Cassie-type state under otherwise comparable impact conditions. The evolution of the wetting state strongly affects droplet freezing time: with decreasing substrate temperature, the shift from the Wenzel to the Cassie wetting regime reduces the droplet-substrate contact area, thereby diminishing thermal transport, resulting in a non-monotonic variation in freezing time, reflecting the coupling between wetting state evolution and nucleation timing. Compared to flat surfaces, droplets on micro-patterned surfaces generally exhibit smaller spreading diameters, with the spreading diameter reduced by at least 20% at high impact velocities, indicating that micro-patterns suppress droplet spreading. Additionally, the effects of droplet impact velocity and initial size were investigated, and a phase diagram depicting droplet impact and freezing regimes as functions of non-dimensional temperature $St$ and the Weber number $We$ was constructed. Overall, this study reveals the influence of micro-patterned surfaces on droplet impact and solidification on supercooled substrates, providing fundamental insight for the development of anti-icing surfaces.

## Acknowledgements

This work was supported by the National Natural Science Foundation of China (Grant No. 52176083 and U25A20191).